\documentclass[%
 reprint,
 amsmath,amssymb,
 aps,
]{revtex4-2}
 
\usepackage{graphicx}
\usepackage{dcolumn}
\usepackage{bm}
\usepackage{natbib}
\setcitestyle{super,sort,compress,comma}

\usepackage[colorlinks]{hyperref}
\usepackage[nameinlink]{cleveref} 
\usepackage{titlesec}
\usepackage{booktabs,tabularx}
\usepackage{enumitem}
\usepackage{url}
\usepackage{microtype}
\usepackage[english=usenglishmax]{hyphsubst}
\usepackage{subfigure}
\usepackage[left]{lineno}
\usepackage{xcolor}
\usepackage{listings}
\usepackage{verbatim}
\usepackage[T1]{fontenc}
\usepackage{enumitem}
\usepackage{makecell}

\usepackage{float}

\makeatletter
\renewcommand{\fnum@figure}{\textbf{\figurename~\thefigure}}
\def\@caption@fignum@sep{ }
\makeatother

\makeatletter
\renewcommand{\fnum@table}{\textbf{\tablename~\thetable}}
\makeatother

\renewcommand{\figurename}{\textbf{Fig.}}
\renewcommand{\tablename}{Tab.}

\newcommand{\ZK}[1]{\textcolor{black}{#1}}
\setlist[enumerate]{itemsep=0.5em, topsep=0.5em, partopsep=0em, parsep=0em}

\hypersetup{
    colorlinks=true,
    linkcolor=blue,     
    citecolor=blue, 
    urlcolor=blue   
}

\smallskip

\titleformat*{\subsubsection}{\scriptsize\bfseries}

\titlespacing*{\section}{0pt}{0.1\baselineskip}{0.2\baselineskip}

\titlespacing*{\section}
{0pt}{2.5ex}{2.5ex}
\titlespacing*{\subsection}
{0pt}{2ex}{2ex}
\titlespacing*{\subsubsection}
{0pt}{2ex}{2ex}

\begin{document}

\title{Amorphous silicon structures generated using a moment tensor potential and the activation relaxation technique \textit{nouveau}}

\author{Karim Zongo$^{1, *}$}
\author{Hao Sun$^{2}$}
\author{Claudiane Ouellet-Plamondon$^{1}$}
\author{Normand Mousseau$^{3}$}
\author{Laurent Karim Béland$^{2, *}$}  

\affiliation{ \vspace{10pt} \textnormal{$^{1}$Département de génie de la construction, École de Technologie Supérieure, Université du Quebec, Montréal, QC, Canada  \\ $^{2}$Department of Mechanical and Materials Engineering, Queen's university, Kingston, ON, Canada \\ $^{3}$Département de physique, Institut Courtois and Regroupement québécois sur les matériaux de pointe, Université de Montréal, Montréal, QC, Canada\vspace{10pt}\\ Correspondence: Karim Zongo (karim.zongo.2@ens.etsmtl.ca) or Laurent Karim Béland (laurent.beland@queensu.ca) }}

\begin{abstract} 
\section*{Abstract}
Preparing realistic atom-scale models of amorphous silicon (a-Si) is a decades-old condensed matter physics challenge. Herein, we combine the Activation Relaxation Technique \textit{nouveau} (ARTn) to a Moment Tensor Potential (MTP) to generate seven a-Si models containing between 216 and 4096 atoms. A thorough analysis of their short-range and medium-range structural properties is performed, alongside assessments of excess energy and mechanical properties. The seven ARTn-MTP models are compared with available experimental data and other high quality a-Si models present in the literature. The seven ARTn-MTP a-Si models are in excellent agreement with available experimental data. Notably, several of our models, including the 216-atom, 512-atom, and 1000-atom a-Si models, exhibit low coordination defects without any traces of crystalline grains. Historically overlooked in previous research, our study underlines the need to assess the validity of the continuous random-network hypothesis for the description of \textit{perfect} amorphous model by characterizing local crystalline environment and to explore the crystallisation process of a-Si through modelling. 
\end{abstract}

\maketitle
\section{Introduction}

Amorphous silicon (a-Si) is used in applications that include photovoltaics, thin-film transistors, battery electrodes, and liquid-crystal displays \cite{street1999technology, carlson1976amorphous, powell1989physics, cui2009crystalline, key2009real}.  Recent advances include its use as coating materials for next-generation gravitational wave detectors \cite{birney2018amorphous}. Additionally, a-Si serves as the canonical disordered solid prototype for the development of modeling methods. 

a-Si \cite{lewis2022fifty,madanchi2024future}  is idealized as a continuous random network (CRN) \cite{zachariasen1932atomic}, characterized by a local atomic environment akin to crystalline silicon \cite{wooten1985computer, vink2001device}. It maintains a local tetrahedral crystalline environment without long-range order. Real a-Si, however, is believed to contain defects associated with its disordered nature at a level that depends on its preparation\cite{roorda1991structural, laaziri1999high, laaziri1999high_xray, levesque2022internal, birney2018amorphous}. These include local coordination defects, medium-range ring defects and two-level systems--the latter contribute to noise in sensitive gravitational wave detectors and cause decoherence in quantum computers \cite{levesque2022internal, birney2018amorphous}. Variations in tetrahedral geometry and bond lengths also significantly affect a-Si's performance. Controlling structural anomalies during preparation poses challenges from both an experimental and a theoretical perspective. Indeed, the structure of a-Si is highly influenced by preparation methods \cite{holmstrom2016dependence}; for instance, chemical vapor deposition \cite{kail2011configurational} and ion implantation \cite{roorda1989calorimetric, van1992evidence, laaziri1999high} can introduce voids and coordination defects. Thermal annealing \cite{roorda1989calorimetric, fortner1989radial, laaziri1999high, haberl2009structural} often refines these imperfections, enhancing material quality. \ZK{Additionally, a recent study \cite{rosset2025signatures} has raised an important issue that questions the definition of the true model of a-Si. This study explored the footprints of paracrystallinity in amorphous silicon using molecular dynamics simulations with a machine learning potential \cite{rosset2025signatures, morrow2022indirect} similar to the one employed herein. As reported by the authors, while optimal models of a-Si are traditionally based on the continuous random network model, some groups have proposed that real a-Si could be better described as 'paracrystalline'. However, since both optimal types of structure conflict with experimental observations \cite{treacy1998paracrystallites,voyles2001structure,treacy2012local,roorda2012comment}, the authors have further investigated the paracrystalline model, concluding that realistic experimental models of a-Si may contain some degree of crystalline grain traces. The observed degree of crystallinity is likely influenced by the preparation method—i.e. MD. Notably, while the degree of crystallinity was examined, its relationship to coordination defects was not. In other words, the central question remains how to balance crystallinity and coordination defects when preparing a-Si. For instance, a high cooling rate in melt-quench simulations will likely lead to a higher number of coordination defects and lower crystallinity, while a low cooling rate will most likely yield fewer coordination defects and higher crystalline-like environments.}

Extensive theoretical research has focused on developing realistic structural models for a-Si \cite{wooten1985computer, barkema1996event, barkema2000high, deringer2018realistic, justo1998interatomic, ishimaru1997generation, wright2013eighty}. Various methods for simulating materials, including a-Si, can be categorized into dynamic and non-dynamic approaches \cite{lewis2022fifty, madanchi2024future}. Dynamic methods, traditionally based on density functional theory (DFT) \cite{sholl2022density} or semiempirical force fields \cite{torrens2012interatomic}, primarily use melt-quench processes. However, these methods face challenges: DFT simulations are limited in system size and quenching rate, and semi-empirical force fields are often compromised by limited accuracy and transferability issues.

Non-dynamic methods include the Reverse Monte Carlo (RMC) approach \cite{mcgreevy1988reverse, mcgreevy1992rmc, gereben2007new, gurman1990reverse, mcgreevy2001reverse}. In RMC, the computer-generated a-Si structure is randomly modified in order for its simulated structure factor and radial distribution function g(r) to match experimental data. RMC's main drawback is that it can yield multiple configurations for the same g(r) --- in other word the solution is degenerate. To overcome this, hybrid RMC methods incorporate force field interactions, ensuring generated configurations are both consistent with experimental data and physically realistic. Variants like the force-enhanced atomic refinement \cite{pandey2015force, igram2018large} and experimentally constrained molecular relaxation \cite{biswas2004inclusion} utilize semi-empirical and DFT interactions, respectively. Other key non-dynamic methods include the Wooten-Winer-Weaire (WWW) bond-switching algorithm \cite{wooten1985computer}, which optimizes atomic configurations by minimizing energy, and the Activation-Relaxation Technique \textit{nouveau} (ARTn) \cite{barkema1996event, malek2000dynamics, mousseau2012activation, jay2022activation}, which combines activation and relaxation steps to discover low-energy structures. ARTn can be paired with various force fields and \textit{ab initio} methods, making it suitable for applications that include relaxation of disordered systems and diffusion processes \cite{mousseau1999exploring, levesque2022internal, ganster2012first, el2007ab, salles2017strain,joly2012optimization,sun2022statistical,sun2023calculation,trochet2020off,beland2011kinetic}.  

One of the key components in both dynamical and non-dynamical methods is the interaction potential. The ongoing challenge of balancing accuracy and computational cost in these methods has led to the development of machine learning interatomic potentials (MLIPs) \cite{behler2007generalized,deringer2019machine,wang2024machine}. MLIPs combine descriptors of the atomic environment, regression methods, and DFT data. Many MLIPs frameworks have been developed, including artificial neural networks \cite{behler2007generalized,khorshidi2016amp}, kernel-based methods \cite{bartok2010gaussian, bartok2018machine}, and linear regression approaches \cite{thompson2015spectral, shapeev2016moment}, among others. For a detailed description of MLIPs, including their implementation and testing, the reader is referred to the existing literature \cite{wang2024machine}. In the case of a-Si, structures have been modeled using artificial neural networks \cite{li2020unified,fan2021neuroevolution,li2019dependence} and Gaussian Approximation Potentials (GAP) \cite{bartok2010gaussian,bartok2018machine}. In initial studies, the GAP potential was applied to a-Si systems containing between 512 and 4,096 atoms \cite{deringer2018realistic}. Later, the GAP model was extended to larger systems of up to 100,000 atoms to investigate the the atomistic mechanisms underlying various structural transitions in disordered silicon \cite{deringer2021origins}. Other MLIPs, such as the Moment Tensor Potential \cite{shapeev2016moment} (MTP), have also been applied to a-Si \cite{zongo2024unified}. In these studies, melt-quench MD simulations were employed to generate a-Si structures. 

A natural next step is to couple ARTn \cite{barkema1996event, malek2000dynamics, mousseau2012activation, jay2022activation} with machine learning potential to generate a-Si models. The objective of this paper is to generate a-Si models using ARTn-ML, which could enhance our understanding of the a-Si model and open new avenues for further research. The manuscript is organized into three main sections. The Methods section outlines the principles of the ARTn method and details our input models, the ARTn-MTP simulation process, and the analytical tools used. The Results and Discussion Section presents findings from our analysis of the configurations generated by ARTn-MTP. Finally, the key insights and discoveries of our study are summarized in the conclusions. 

\section{Methods}
\subsection{Activation Relaxation Technique \textit{nouveau}}

ARTn \cite{barkema1996event, malek2000dynamics} is a computational method used in the study of activated processes, particularly in the context of diffusion and transitions between states in condensed matter systems. ARTn is designed to simulate rare events that signify substantial changes in the microscopic structure of atomic systems, based on the efficient exploration of potential energy landscapes. ARTn employs an event-driven approach in which the system evolves by transitioning between different states or configurations. The concept is based on the understanding that within the configurational space, there is a distinct set of points representing saddle points and adjacent minima in the potential energy surface of the system. As the name implies, ARTn can be summarized as a two-step process \cite{barkema1996event}: activation and relaxation. In the activation step, starting from a local energy minimum, the configuration is pushed toward a nearby first-order saddle point. In the relaxation step, the system is driven over the saddle point and then relaxed to a new energy minimum. 
The new configuration is accepted based on the Metropolis acceptance probability, which is given by $ P_{\text{accept}} = \min \left(1, \exp \left(-\frac{\Delta E}{kT}\right) \right)$, where T is the fictitious temperature, set to 0.25 eV for our simulation, k is the Boltzmann constant, and $\Delta E$ is the energy difference between the old and new configurations \cite{barkema1996event, RN210,barkema2000high}. The search is restricted to paths that include only saddle points. The ARTn method has demonstrated its value in revealing rare events that conventional simulation methods may struggle to capture due to the slow dynamics and relaxation inherent in these systems. Detailed descriptions of the method can be found elsewhere \cite{barkema1996event,malek2000dynamics,mousseau2012activation,jay2020,gunde2024}.

\subsection{A moment tensor potential for silicon}

In practical terms, ARTn serves as a saddle-point search engine, employing an interaction potential as both a force and energy calculator. The simulation's accuracy is directly influenced by the selection of the force field. Previous studies have relied on semi-empirical force field \cite{ganster2012first,barkema1996event,mousseau1998traveling,levesque2022internal,eliassen2019atomistic} and ab initio methods \cite{machado2011optimized, beland2014strain,salles2017strain}. Here we employ a machine learning-based forcefield. This potential was developed to jointly describe Si, O, and SiO$_2$ systems \cite{zongo2024unified}, and is an improvement over a previously reported Si MTP \cite{zongo2022first}. This potential is based on the MTP framework \cite{shapeev2016moment, novikov2020mlip}. \ZK{The MTP potential was trained using data obtained from DFT calculations with the generalized gradient approximation (GGA-PBE) as the exchange-correlation functional \cite{perdew1996generalized}. The MTP training set did not include amorphous silicon (a-Si) configurations. As explained in Ref. \cite{zongo2024unified}, the MTP was validated against ab initio MD by performing a rapid melt-and-quench. Fast cooling (–8 ps) was followed by equilibration (10 ps) using either ab initio MD or MTP-MD. Both ab initio MD and MTP-MD led to a-Si models with very similar structural properties while repeating the same protocol with semi-empirical potentials led to substantially different results. The simulations presented herein can be seen as further validation and testing of this MTP.} A detailed description of the development and implementation of this MTP can be found in our previous work \cite{zongo2024unified,zongo2022first}.

\subsection{Nomenclature of the models}

\ZK{To name our models, including those from the literature, we used several techniques. Some models were named based on their creation procedure, while others were named according to the simulation method and the year the model was generated. For our models, we used the following naming conventions. Five of our models were named 216-R1, 216-R2, 512-R1, 512-R2, and 1000-R. Here, R stands for randomly generated initial structure, meaning the simulation box was randomly filled with atoms. When there are multiple models with the same number of atoms, the first model is named R1, and the second is named R2. If there is only one model for a given number of atoms, it is simply named R, as in 1000-R. One of our models was initially randomly filled and then pre-relaxed with MD simulation, so it was named 4096-R-MD. The models named 216-MD1, 216-MD2, 512-MD1, 512-MD2, 1000-MD, and 4096-MD refer to configurations that were entirely simulated using Molecular Dynamics (MD) with the ML (MTP) potential. The initial configurations are the same as those used for 216-R1, 216-R2, 512-R1, 512-R2, 1000-R, and 4096-R-MD, meaning the initial setups were randomly generated as described earlier. We found these input configurations to be a good starting point, as opposed to starting from a crystalline state and melting it. Models from the literature retain their original preparation methods, as indicated in the last column of Table I. For example, 216-FEAR indicates that the a-Si model was prepared using Force-enhanced atomic refinement, and 100k-GAP18 refers to a model with 100,000 atoms of a-Si, prepared using the Gaussian Approximation Potential (GAP) developed in 2018 for silicon materials.}

\begin{table*} 
\begin{tabular}{lccc}
\hline
Models name & Input configurations for a-Si preparation & N atoms & a-Si preparation methods\\
 \hline
216-R1  & Randomly filled & 216 & ARTn runs using MTP potential\\
216-R2  & Randomly filled & 216 & ARTn runs using MTP potential\\
216-MD1  & Randomly filled & 216 & MD simulation using MTP potential\\
216-MD2  & Randomly filled & 216 & MD simulation using MTP potential\\
216-FEAR  & Randomly filled & 216 & Force-enhanced atomic refinement \cite{igram2018large} \\
512-R1  & Randomly filled & 512 & ARTn runs using MTP potential\\
512-R2   & Randomly filled & 512 & ARTn runs using MTP potential\\ 
512-MD1  & Randomly filled & 512 & MD simulation using MTP potential\\
512-MD2  & Randomly filled & 512 & MD simulation using MTP potential\\
512-RMC & Randomly filled & 512  & Reverse monte carlo simulation \cite{cliffe2017structural,deringer2018realistic}\\
512-INV & Randomly filled & 512  & Invariant environment refinement technique \cite{cliffe2017structural,deringer2018realistic}\\
512-FEAR & Randomly filled & 512  & Force-enhanced atomic refinement \cite{igram2018large} \\
512-WWW & Randomly filled &  512 & Bond switching algorithm \cite{cliffe2017structural,deringer2018realistic}\\
1000-R & Randomly filled & 1000  & ARTn runs using MTP potential \\
1000-MD & Randomly filled & 1000  & MD simulation using MTP potential \\
4096-R-MD & Randomly filled + MD runs & 4096  & ARTn runs using MTP potential \\
4096-MD & Randomly filled + MD runs & 4096  & MD simulation using MTP potential \\
4096-WWW & Randomly filled &  4096 & Bond switching algorithm \cite{igram2018large} \\
100k-GAP18 & Unknown & 100000 & MD simulation using GAP potential \cite{deringer2021origins}\\
1000-aSi-SW & Amorphous configuration \cite{barkema1996event,malek2000dynamics,machado2011optimized} & 1000 & ARTn runs using MTP potential\\
\hline
\end{tabular} 
\caption{\label{tab:prepam} A summary of the models discussed in this study. The preparation methods for the input configurations, the number of atoms in each model, and the preparation methods for the final a-Si models are provided. Models without a reference were prepared for this work following the protocol discussed in the methods section.}
\end{table*}
\begin{figure*}
    \centering
    \subfigure[]{\includegraphics[width=0.60\textwidth]{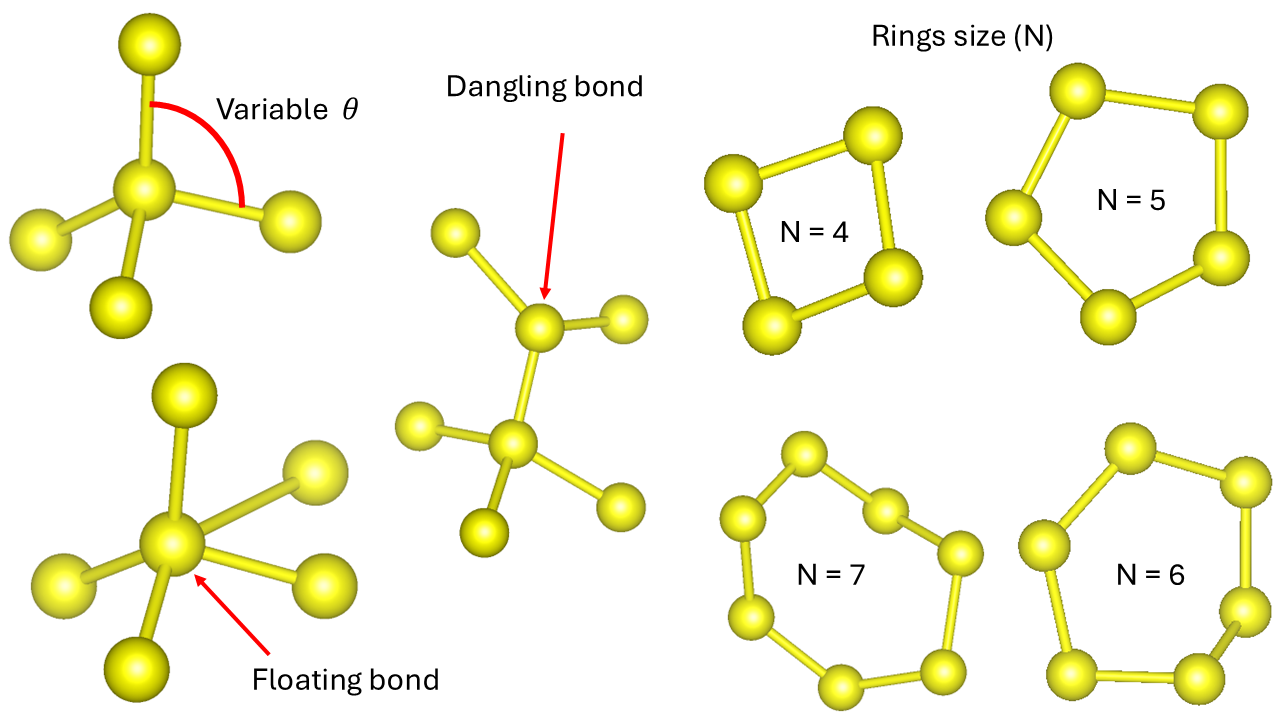}} 
    \subfigure[]{\includegraphics[width=0.37\textwidth]{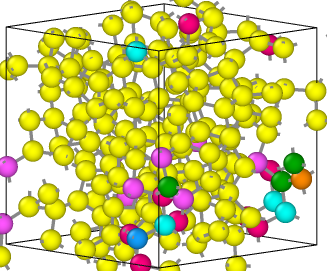}} 
    \caption{Key structural  elements characterizing a-Si: (a) bond angle ($\theta$), coordination defects such as dangling bonds (3-fold bonded atoms) and floating bonds (5-fold bonded atoms), and ring structures, including 4-, 5-, 6-, and 7-membered rings. (b) A 100\% 4-fold fully coordinated amorphous structure comprising 216 atoms, generated by ARTn-MTP simulation , with 14.4 \% of atoms in a crystalline environment. The atoms shown in yellow within the box are in an amorphous environment, while those colored differently are in a crystalline environment, specifically cubic or hexagonal diamond.}
    \label{fig:ims}
\end{figure*}
\subsection{Protocol for input models preparation and simulation details}

We prepared models containing 216, 512, 1000, and 4096 atoms. For the model with 216 atoms, we used two different simulation boxes corresponding to densities of 2.20 g/cm³ (216-R1) and 2.28 g/cm³ (216-R2), respectively. These two boxes were randomly filled with an overlap distance (initial atomic separation) of 2.3 Å for 2.20 g/cm³ and 2 Å for 2.28 g/cm³. For the 512-atom model, we employed two simulation boxes randomly filled with atoms, using the same densities as those applied to the 216-atom model. The first 512-atom model had a density of 2.20 g/cm³ (512-R1) with an initial overlap of 2.3 Å. The second 512-atom model (512-R2) along with the 1000-atom model (1000-R) were randomly filled with a density of 2.28 g/cm³ and an initial overlap of 2 Å. The search for events during ARTn-MTP simulations was initiated using configurations that were already at local minima. Herein, the ARTn simulations begins by applying random displacements to the chosen central atom and its neighbors, based on a user-defined local cut-off radius. Once the displacements are introduced, the algorithm searches for convergence to a saddle point, guided by the lowest curvature. After locating the saddle point, the minimization process begins to find the new minimum. The minimization is carried out using either the Steepest Descent (SD) or the Fast Inertial Relaxation Engine (FIRE) algorithm. In all ARTn-MTP simulations, we used a fictitious temperature of 0.25 eV in the Metropolis accept-reject criterion.\ZK{The fictitious temperature is not a thermodynamic temperature of our system as we use ARTn here was a way to minimize total energy through activated events. Since no thermodynamical ensemble is therefore defined, this parameter serves as an artificial tool to aid in sampling configurations and achieving better convergence in the simulation.} Each simulation started with event searches initiated with a local cut-off radius of 3 Å, which the algorithm used to identify the region around the selected central atom which will be deformed. \ZK{The cut-off radius defines a spherical region around the central atom, within which both the selected central atom and its neighboring atoms are displaced during each event.} This cut-off radius was progressively increased to a maximum cut-off  5.5 Å in order to expand the local zone of initial deformaton when the ARTn moves no longer led to energy decrease. The coordination defects level was monitored during the simulations. For coordination defect levels above 5 \%, the full list of atoms was used as potential central atoms for ARTn moves. For defect levels at or below 5 \%, only atoms with a non-four-fold coordination were allowed to serve as the central atom in ARTn moves. \ZK{For comparison purposes, we used the same initial (input) configurations as those for the ARTn-MTP simulations to run MD simulations with the MTP potential. For each MD model—216-MD1, 216-MD2, 512-MD1, 512-MD2, 1000-MD, and 4096-MD—the configuration was first heated to 2500 K over 30 ps, followed by equilibration at this temperature for 1 ns. The system was then cooled at a rate of 1 K/ps, and finally, the configuration was equilibrated at 300 K for 150 ps using the isothermal-isobaric ensemble (NPT).}

\subsection{Structural analysis}

\ZK{We use a set of criteria to assess the quality of our amorphous structures, based on short-range order, coordination defects, intermediate-range order, and the presence of crystalline grains, as illustrated in Fig. \ref{fig:ims} a)}. Understanding these different structural aspects is crucial in the study of materials properties, as they influence the physical, chemical, and mechanical behavior of materials. We conduct coordination analysis using OVITO \cite{stukowski2009visualization} with a bond-length cutoff of 2.85 Å. The analysis of the ring size distribution, providing insights into intermediate-range order, is performed using the R.I.N.G.S code \cite{{le2010ring}} based on the King criteria. We employed the ISAACS code \cite{le2010isaacs} for the calculation of the structure factor while the radial distribution function and angular distribution function were determined using LAMMPS \cite{thompson2022lammps} through pseudo simulations.

\ZK{We also quantify the crystallinity in the models, as indicated by the colored atoms (excluding the yellow atoms) in Fig. \ref{fig:ims} b}. Here, crystallity refers to environments that are cubic or hexagonal diamond to second neighbour, all atoms at the center of this environment or within the first- and second-neighbour shell of a perfectly atom are considered as having crystallinity. We use method developed by  Ref.~\cite{maras2016} as implemented in OVITO~\cite{stukowski2009visualization}. \ZK{To verify the crystalline content of our models, we used Common Neighbor Analysis (CNA) in OVITO, which offers four operation modes. Two are relevant for amorphous systems: the conventional CNA with a fixed cutoff and the Adaptive CNA with a variable cutoff. The conventional CNA is better suited for crystalline systems, while the Adaptive CNA is designed for multiphase systems, which is more relevant for amorphous systems with crystalline seeds. Both methods were applied to our models, but they are limited to lattice systems other than diamond and hexagonal diamond. Therefore, we then used the "Identify Diamond Structure" mode in OVITO, an extension of Adaptive CNA, which automatically determines the optimal cutoff radius for each particle. Finally, we varied the cutoff in conventional CNA but observed no changes in crystallinity.}

\section{Results and discussion}

\begin{table*} 
\caption{\label{tab:Defects} Structural characteristics of ARTn-MTP generated amorphous models. This includes density, the positions of the first (r$_{1}$) and second (r$_{2}$) peaks of the RDF, average coordination number, average bond angle (with variability indicated in parentheses), and the percentages of 3-fold, 4-fold, and 5-fold coordinated atoms in each mode. A comparison is provided with previously published models. Bold fonts are used to highlight models where fewer than 3 \% of atoms were mis-coordinated. Crystallinity is defined as discussed in the Methods section. Cells are left empty when information is not available for models from the literature. }
\begin{tabular}{lccccccccc}
\hline
Model & density& r$_1$ & r$_2$ & Coordination number & $\langle \theta \rangle$ ($\sigma$) & 3-fold (\%) & 5-fold (\%) & 4-fold (\%) & Crystallinity (\%) \\
 \hline
216-R1 & 2.20  & 2.36 & 3.77 & \textbf{3.990} & 109.20 (9.95) & 0.93 & 0.00 & \textbf{99.07} & 0.00 \\
216-R2 & 2.28  & 2.36 & 3.70 & \textbf{4.000} & 109.19 (10.02) & 0.00 & 0.00 & \textbf{100.00} &14.40 \\
216-MD1 & 2.25  & 2.37 & 3.83 & \textbf{4.018} & 109.08 (11.51) & 0.00 & 1.85 & \textbf{98.15} &0.00 \\
216-MD2 & 2.26  & 2.37 & 3.83 & \textbf{4.000} & 109.05 (10.60) & 0.93 & 0.93 & \textbf{98.14} &7.90 \\
216-FEAR \cite{igram2018large}  & 2.33  & 2.36 & 3.81 & 4.028 & 108.52 (15.59) & 1.39 & 4.17 & 94.44 & \\
512-R1 & 2.20  & 2.36 & 3.90 & 4.008 &109.11 (11.39) & 0.59 & 1.37 & \textbf{98.04} & 0.00\\
512-R2 & 2.28  & 2.36 & 3.90 & 4.015 & 108.99 (11.76) &  0.19 & 1.76 & \textbf{98.04}& 0.00\\
512-MD1 & 2.24  & 2.37 & 3.83 & 4.004 & 109.09 (11.10) &  0.78 & 1.17 & \textbf{98.05} & 3.30\\
512-MD2 & 2.24  & 2.37 & 3.90 & 3.996 & 109.13 (10.44) &  1.37 & 0.98 & 97.65 & 3.30\\
512-RMC \cite{cliffe2017structural,deringer2018realistic} & 2.30  & 2.34 & 3.66 & 4.039 & 108.57 (14.16) & 1.56 & 5.47 & 92.96 & 3.30 \\ 
512-INV \cite{cliffe2017structural,deringer2018realistic}& 2.31  & 2.34 & 3.86 & 4.054 & 108.56 (14.47) & 1.95 & 7.42  & 90.62 & 0.00\\
512-WWW \cite{cliffe2017structural,deringer2018realistic}& 2.27  & 2.34 & 3.94 & 4.008& 109.05 (10.36) & 0.19 & 0.98 & \textbf{98.82} & 10.00\\  
512-FEAR \cite{igram2018large} & 2.33  & 2.35 & 3.82 & 4.008 &  & 1.17 & 2.73 & 95.90 &  \\
512-GAP18 \cite{deringer2018realistic} & 2.27  & 2.36 & 3.76 & 4.004 & 109.20 (9.60) & 0.58 & 0.98 & \textbf{98.44} & 6.40 \\
1000-R & 2.28  & 2.37 & 3.76 & 4.020 & 108.94 (12.31) & 0.60 & 2.60 & 97.20  & 0.00 \\
1000-MD & 2.25  & 2.37 & 3.83 & 4.016 & 109.98 (11.25) & 0.70 & 2.30 & 97.00  & 13.30 \\
4096-R-MD & 2.28  & 2.37 & 3.83 & 4.012 & 109.14 (10.65) & 0.51 & 1.44 & \textbf{98.05} & 6.40 \\
4096-MD & 2.25  & 2.37 & 3.90 & 4.009 & 109.06 (11.04) & 0.78 & 1.66 & 97.56 & 11.70 \\
4096-WWW \cite{igram2018large} & 2.33  & 2.36 & 3.78 & 4.004 &  & 0.05 & 0.49  & \textbf{99.46} &  \\
100k-GAP18 \cite{deringer2021origins} & 2.27  & 2.36  & 3.83 & 4.003 & 109.17 (9.99) & 0.69 & 0.95  & \textbf{98.35} & 6.5\\ 
\hline
\end{tabular}    
\end{table*}

\begin{figure*}
\centering
    \includegraphics[width=18cm]{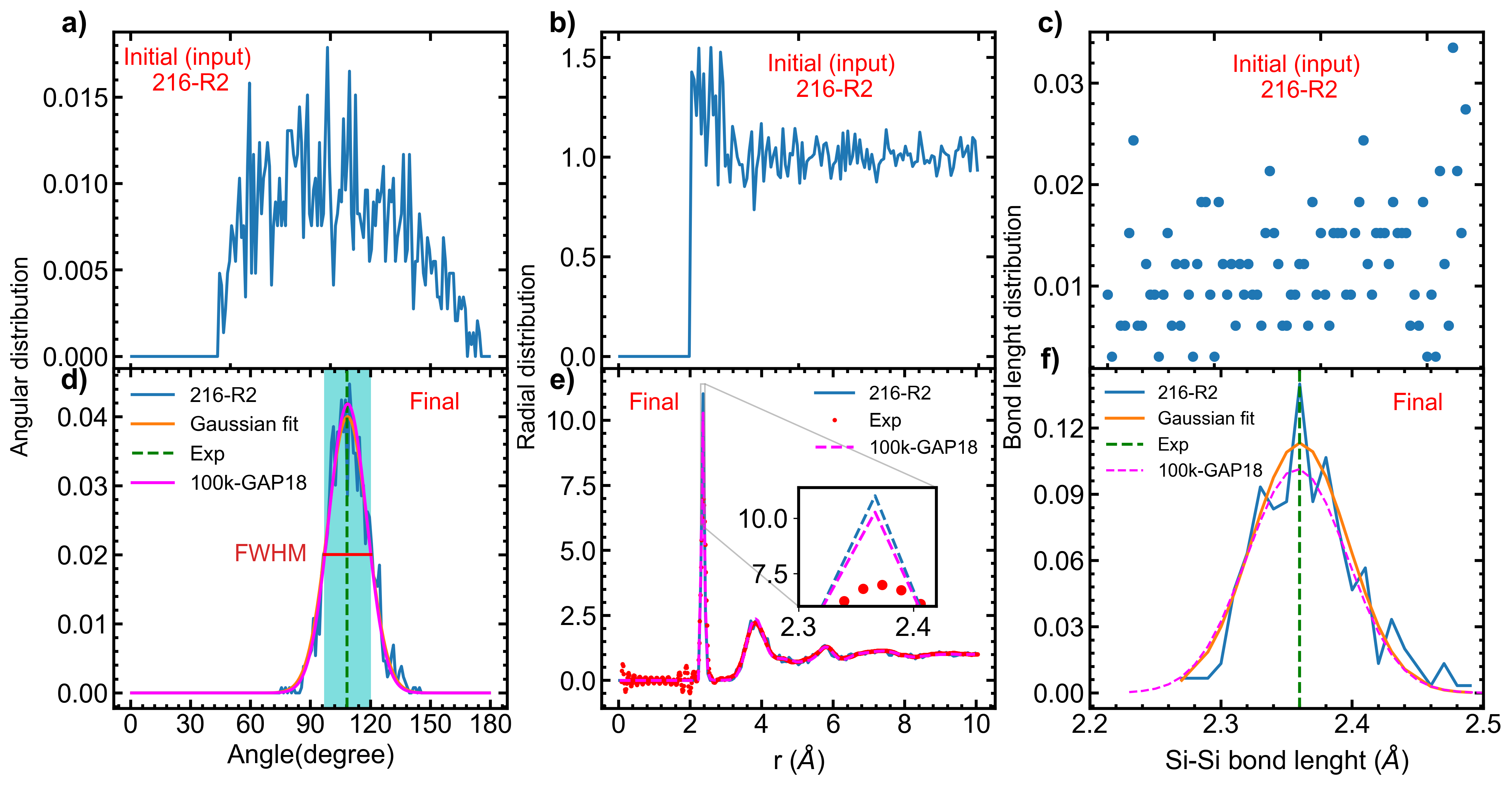}
    \caption{\label{fig:Struc216} Comparison of the local structural properties of input configurations (top panels: a, b, and c) with the final configuration (100\% coordinated) (bottom panels: d, e, and f) generated using the ARTn-MTP coupling scheme. The structural properties analyzed include angular distribution, radial distribution, and bond length distribution for a system comprising 216 atoms. Experimental data and data from the 100k-GAP18 model \cite{deringer2021origins} are included to assess the quality of our 100\% 4-fold coordinated 216-atom a-Si model
    }
\end{figure*}

Models of size 216 atoms, 512 atoms, 1000 atoms and 4096 atoms were generated as detailed in table \ref{tab:prepam}. We analyze these configurations with respect to short-range order, medium-range order, as well as mechanical properties. First, we would like to emphasize that, the input configurations exhibit a higher prevalence of coordination defects than the final configurations: approximately 70\% for the 216-R1, 216-R2, and 512-R1 models, compared to around 30\% for the 512-R2 and 1000-R models.

In all ARTn-MTP a-Si models, the final percentage of defective atoms remained below 2\%, with the exception of the 1000-R and 4096-R-MD models, which exhibit a final defect percentage of approximately 3\%. The initial 4-fold coordination number is approximately 30\% for the 216-R1, 216-R2, and 512-R1 models, and around 68\% for the 512-R2 and 1000-R models. In contrast, the final configurations display markedly higher 4-fold coordination numbers (Table~\ref{tab:Defects}), with values of 99.07\%, 100\%, 98.04\%, 98.04\%, 97.20\%, and 98.05\% for the 216-R1, 216-R2, 512-R1, 512-R2, 1000-R and 4096-R-MD models, respectively. No atoms with fewer than 3 neighbors or more than 5 neighbors were observed.

To further contextualize the number of coordination defects in the ARTn-MTP a-Si models, we present a comparison in Table~\ref{tab:Defects}, which includes data from MD simulation and the literature as well. First, all the models presented show no atoms with fewer than 3 neighbors or more than 5 neighbors. Secondly, all of the ARTn-MTP generated models exhibit fewer floating and dangling bonds compared to all other methods, except the WWW model and 100k-GAP18 model. Notably, the 216-R2 model exhibit 100\% four-fold coordination. To our knowledge, no other a-Si models exhibit 100\% four-fold coordination, except for some WWW models generated with a Keating-type potential \cite{pandey2016inversion,barkema2000high}

The last column of Table~\ref{tab:Defects} presents the fraction of atoms in a crystalline environment (cubic or hexagonal diamond). As shown, most of the generated a-Si models, particularly those from MD simulations, contain some proportion of crystallinity. Although WWW is known to produce large a-Si models without coordination defects, the WWW a-Si models still exhibit traces of crystallinity. We identified four models—216-R1, 512-R1, 512-R2, and 1000-R—that are free from crystallinity. Furthermore, as indicated in Table ~\ref{tab:DefectsSW}, both the 1000-aSi-SW initial and 1000-aSi-SW final configurations also contain no crystallinity. It is important to note that all four of our a-Si models without crystallinity, including the 1000-aSi-SW initial model, were generated starting from a random state without any prior MD prerun, meaning their boxes were initially filled randomly. 
From Table~\ref{tab:Defects}, we note three potential trends. (i)
this fraction does not seem to depend much on the system size. (ii) Starting ARTn from a random state seems to lead to a less crystalline state than starting from a melt-and-quench state. (iii) Fewer coordination defects seem to favor a higher crystallinity content. Now that, with reliable ML potentials, we can generate similar quality models using different pathways (non-dynamic melt-and-quench methods), we can start examining the link between local crystallinity, defects and the ideal continuous-random networks. This can lead to understanding why, for example, amorphous silicon cannot be produced experimentally
from the liquid phase and crystallizes well below melting as well as whether the CRN model corresponds really to ideal a-Si. The ML potential \cite{zongo2024unified,zongo2022first}, which leads to high-quality models, makes further study on this question now possible.

\begin{figure}
\centering
    \includegraphics[width=8.5cm]{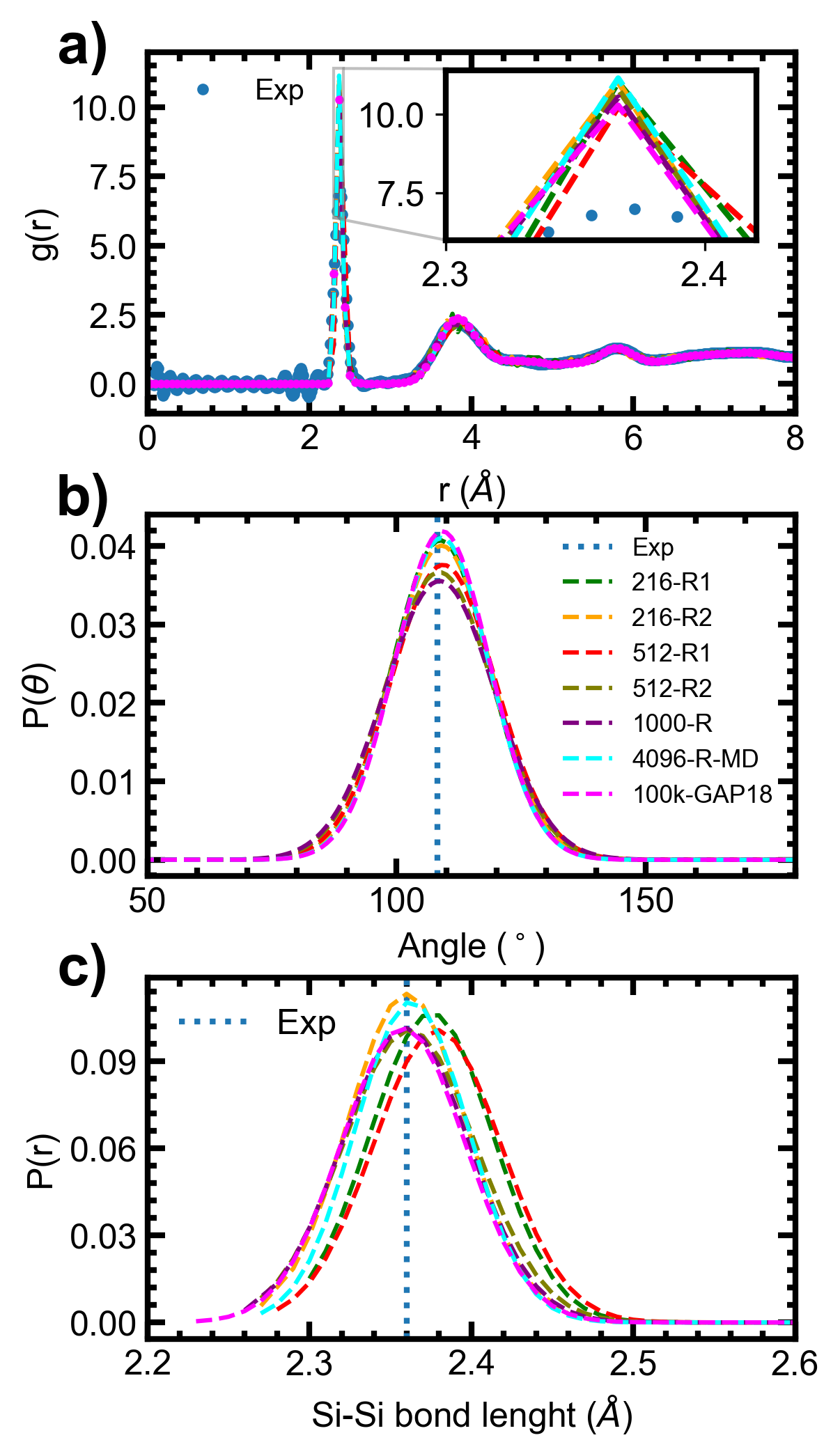}
    \caption{\label{fig:Totalc} Local structural properties of ARTn-MTP generated a-Si models: (a) radial distribution, (b) angular distribution, and (c) bond length distributions.  Also shown are the properties obtained from experimental data and the 100k-GAP18 atom simulations.}
\end{figure}

\begin{figure}
\centering
    \includegraphics[width=8.5cm]{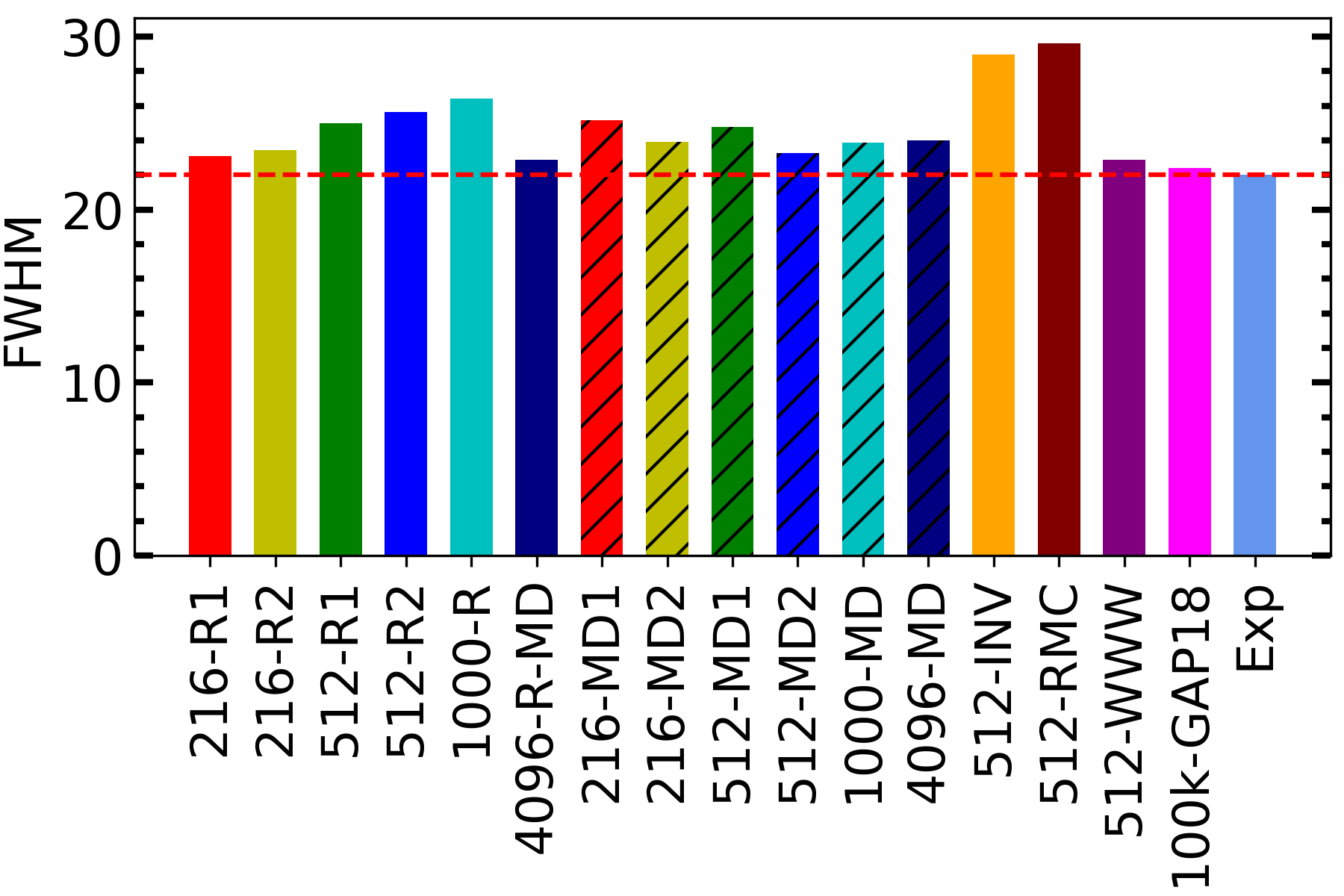}
    \caption{\label{fig:thear} FWHM of the bond-angle distribution. The FWHM values are computed by fitting a Gaussian distribution to the bond angle distributions. The bars are hatched for our purely MD-generated a-Si models. The red line denotes the experimental value of FWHM \cite{fortner1989radial}.}
\end{figure}

\subsection{ Short-range order structural analysis of the models}

Fig. \ref{fig:Struc216} illustrates the structural properties of the 216-R2 models, including the angular distribution function, radial distribution function, and bond length distribution. Panels a, b, and c display these properties for the randomly packed input configuration (216-R2), while panels d, e, and f show the properties of the final configurations (216-R2) produced by ARTn-MTP. The final 216-R2 configuration is 100\% fully coordinated, with no coordination defects present. As shown in Fig. \ref{fig:Struc216} a, b, and c, the data for the input configurations are randomly distributed and widespread. In contrast, the data for the final configurations, depicted in Fig. \ref{fig:Struc216} d, e, and f, follow well-defined distributions. The shape of the radial distribution function matches the experimental data, though the peak is somewhat overestimated. This indicates that the Si-Si pair distances in our model are widely distributed around the experimental pair separation distance.  The bond angle distribution function is centered around the tetrahedral value of 109.5° with a spread of 9.96°. A Gaussian fit to the data reveals a full width at half-maximum (FWHM) of 23.45°. The half width at half-maximum (HWHM) is 11.73°, which is similar to the experimental HWHM value of 11° \cite{fortner1989radial}. For the bond length, the data for the input configuration (Fig. \ref{fig:Struc216} c are widespread. In contrast, for the final configuration (Fig. \ref{fig:Struc216} f, the bond lengths follow a Gaussian distribution centered around the experimental value of 2.35 Å. The mean bond length is 2.36 Å, with a spread of 0.03 Å.  The angular distribution function, radial distribution function, and bond length distribution of the final model (216-R2) were compared with those of the 100k-GAP18 configuration, which contains 100,000 atoms. As shown in Fig. \ref{fig:Struc216} d, e, and f, these properties align well with those of the larger 100k-GAP18 model. This demonstrates the high quality of our fully coordinated a-Si model. Beyond the analysis of the fully coordinated configuration, we also explore the short-range order structure in our other models. The detailed results are presented in Fig. \ref{fig:Totalc} and Tab. \ref{tab:Defects}. Once again, the structural properties—including the radial distribution function (Fig. \ref{fig:Totalc} a), bond angle distribution function (Fig. \ref{fig:Totalc} b), and bond length distribution (Fig. \ref{fig:Totalc} c)—closely match experimental data and those of 100k-GAP18 models. The data for all the models presented in Fig. \ref{fig:Totalc} show similar behavior, with the exception of the Si-Si bond length distribution. Notably, the bond lengths for two of our models—216-R1 (in green) and 512-R1 (in red)—are shifted. These two models have a density of 2.20 g/cm³, which is lower than that of the other models, which is 2.28 g/cm³. Since amorphous silicon features local tetrahedral bonding, we quantify the deviation from ideal tetrahedral geometry by calculating the FWHM for all models included in this study, as presented in Fig. \ref{fig:thear}. The FWHM values for all our models closely align with the experimental data, as well as with the WWW model and the 100k-GAP18 models. 

\subsection{ Medium-range order structural analysis of the models}

\begin{figure}
\centering
    \includegraphics[width=8.5cm]{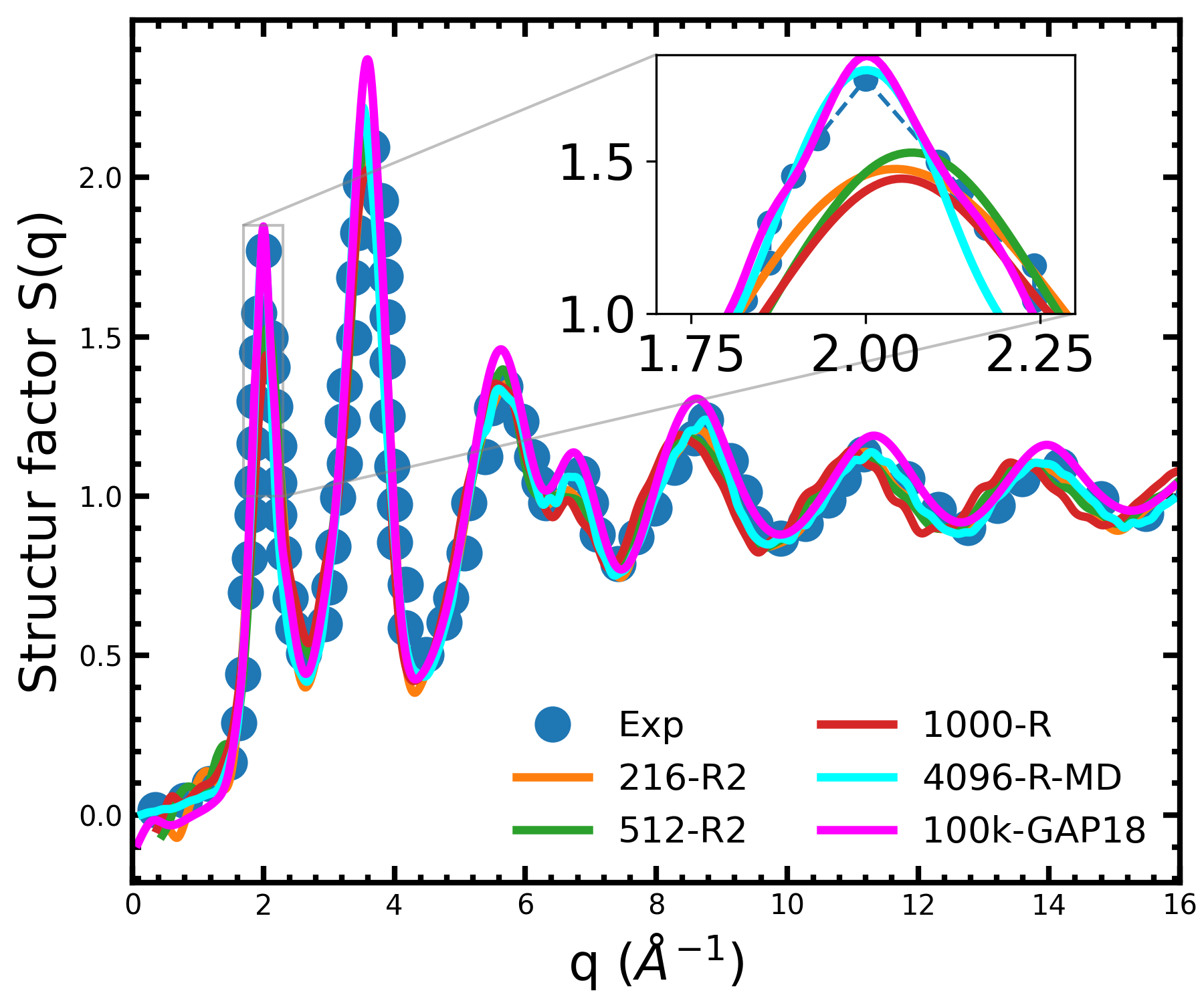}
    \caption{\label{fig:SQcomp} Structure factor for ARTn-MTP generated models compared with that obtained from experimental data from \cite{laaziri1999high} and 100k-GAP18 atom simulations.}
\end{figure}

\begin{figure*}
\centering
    \includegraphics[width=17cm]{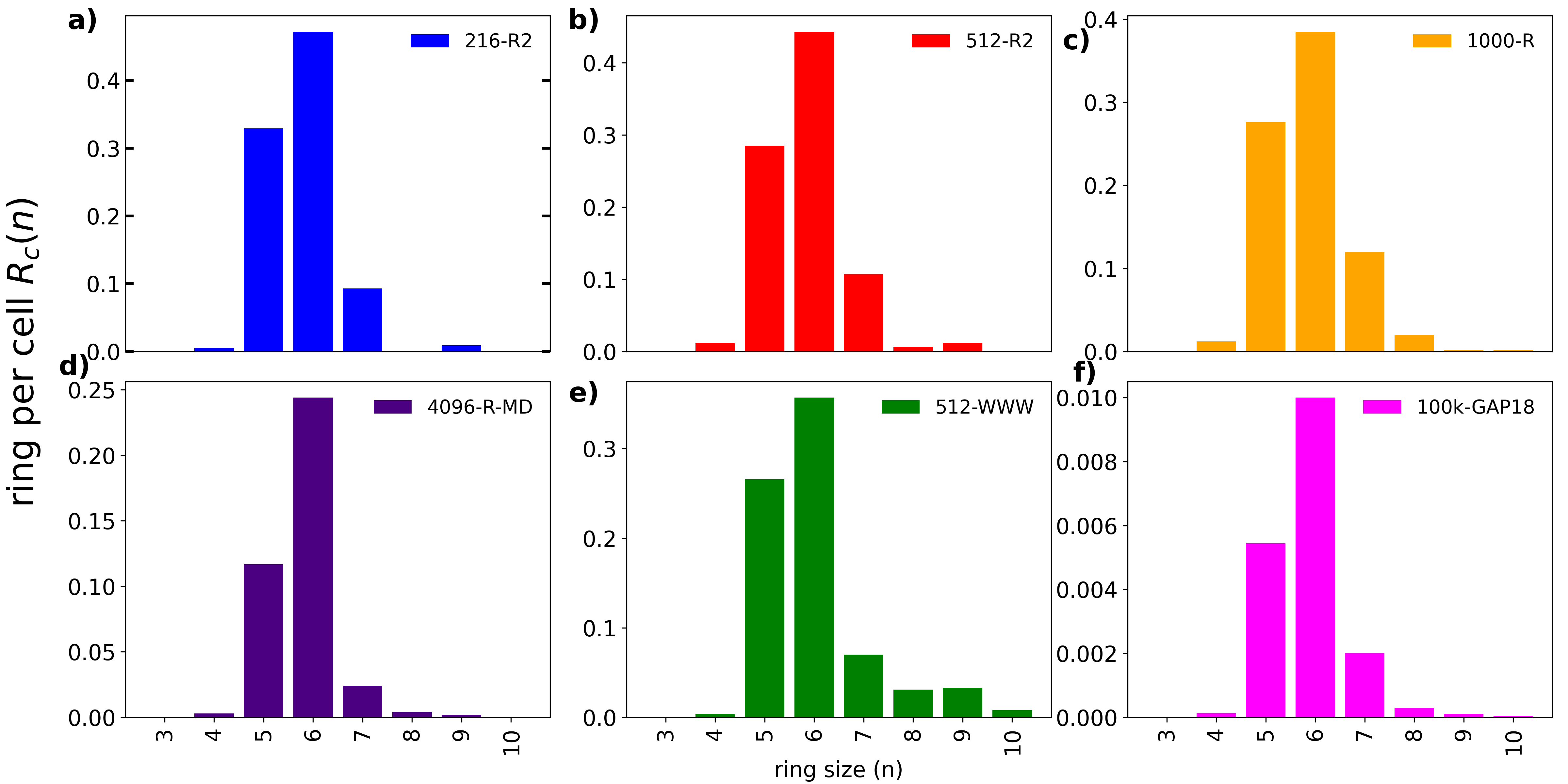} 
    \caption{\label{fig:rings} Ring statistics for ARTn-MTP generated models as well as those derived from the 512-WWW model and 100k-GAP18 atom models. \ZK{The total ring count is normalized by the number of atoms in the model.}}
\end{figure*}

We further analyze the ARTn-MTP a-Si models at intermediate length scales using the structure factor and ring statistics.
Fig. \ref{fig:SQcomp} illustrates the static structure factor of our fully coordinated 216-R2 a-Si model, alongside experimental data and the 100k-GAP18 model. Additionally, the structure factors for the 512-R2, 1000-R, and 4096-R-MD models are included to provide a comprehensive comparison with both the experimental \( S(q) \) and the \( S(q) \) of the 100k-GAP18 models. As shown, the structure factors of all our models align well with the experimental data, with the exception of the first peak at \( S(q) \) near \( q = 2 \, \text{Å}^{-1} \), which is underestimated by the 216-R2, 512-R2, and 1000-R models. This is consistent with other results from the literature\cite{{igram2018large},{deringer2018realistic}} and is attributed to the system size rather than a specific method. Thus, the 4096-R-MD model, which consists of 4096 atoms, and the 100k-GAP18 model, comprising 100,000 atoms, both exhibit a strong match with the first peak of the experimental \( S(q) \). The first peak, known as the first sharp diffraction peak, is typically underestimated in systems with fewer than a thousand atoms\cite{{igram2018large},{deringer2018realistic}}. The height of this peak serves as a measure of structural ordering. \ZK{While the low-k part of structure factor, defined in the reciprocal space, is sensitive to long-range properties and the box size, the real-space structural properties are less affected by the box size. The ring size distribution remains nearly the same as the system size increases, with a clear preference for rings of size 5, 6, and 7, as shown in Fig. \ref{fig:rings}. The bulk modulus for all of our models is distributed around a mean value of 77 GPa. The bond length distribution (Fig. \ref{fig:Totalc} c) is also insensitive to the system size. However, there is still a need to simulate larger systems, as the number of independent environments in a-Si is not yet fully understood. The larger the system, the more diverse local environments can be probed, providing a better understanding of a-Si.}

Ring statistics are illustrated in Fig. \ref{fig:rings}. Crystalline silicon, characterized by its diamond-type structure, primarily features 6-membered rings, while amorphous silicon tends to favor rings of sizes 5, 6, and 7. Rings with fewer than 5 members or more than 7 are typically regarded as ring defects. Notably, no 3-membered rings are observed in any of the models. Additionally, very few 4-membered rings, which are known to introduce structural stress, were recorded across all models. As shown in Fig. \ref{fig:rings} a-f, each model predominantly contains 5, 6, and 7-membered rings, consistent with existing literature. These ring sizes are energetically favorable for a-Si. Furthermore, all models exhibit a few larger ring defects with sizes greater than 7, as detailed in Fig. \ref{fig:rings}. 

\subsection{ Refinement of ART-SW optimized model: 1000-aSi-SW }
 \begin{figure*}
\centering
    \includegraphics[width=18cm]{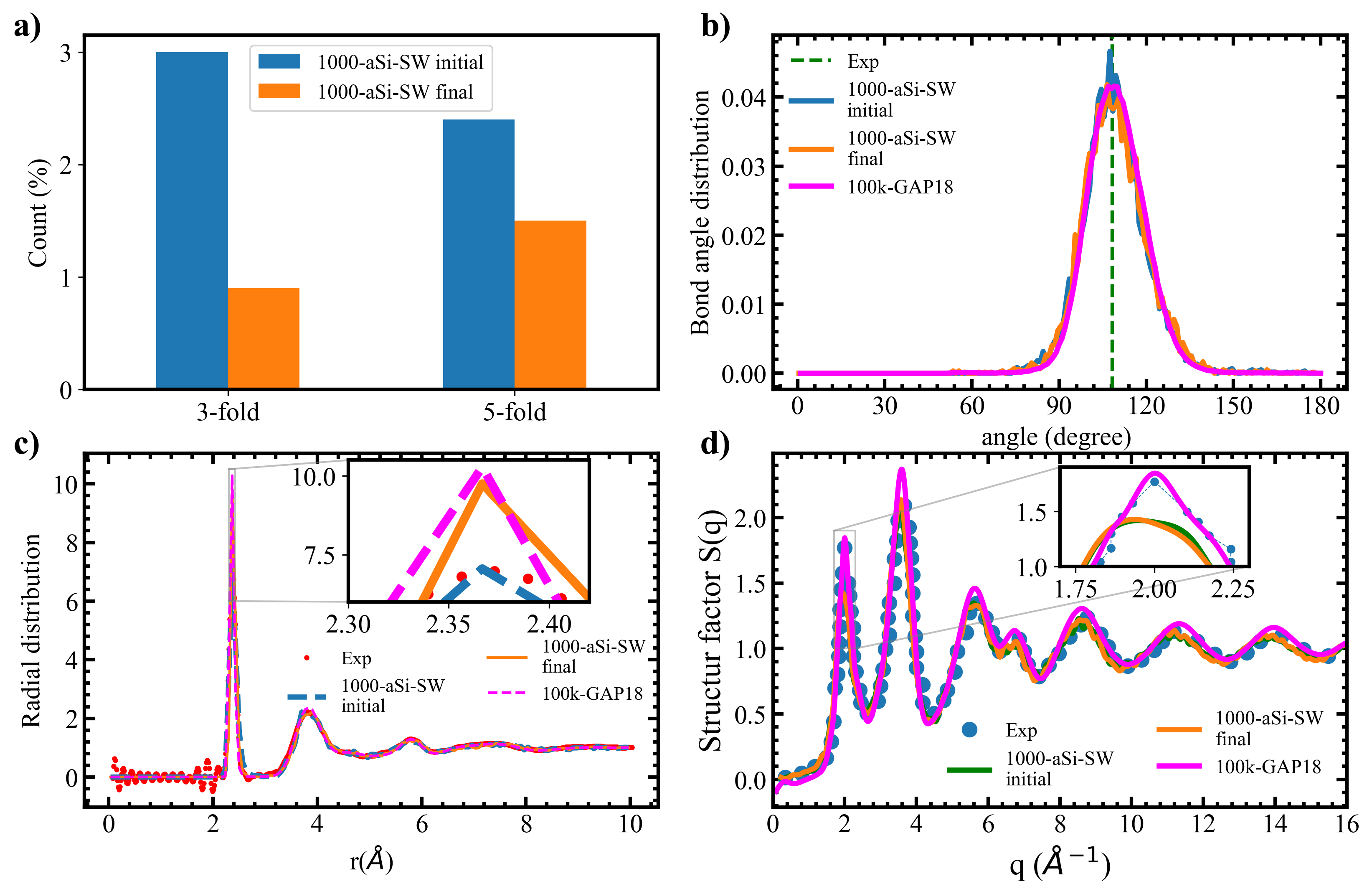}
    \caption{\label{fig:swconf} Local structural properties of ARTn-MTP refined configurations (1000-aSi-SW final) compared to ART-SW optimized configurations (1000-aSi-SW initial). a) Initial and final counts of 3-fold and 5-fold defects, b) bond angle distribution function, c) radial distribution function, and d) structure factor.  For figures (b), (c), and (d), experimental data \cite{laaziri1999high} and results from the 100k-GAP-18 atom model are included to assess the quality of both the initial and final configurations of the 1000-aSi-SW model. The ART-SW optimized configurations served as the input for the ARTn-MTP refined configurations.}
\end{figure*}

As we first test the capability of ARTn coupled with MLIP to model amorphous systems, it is essential to explore several scenarios. A key consideration is the starting input configuration (see Table \ref{tab:prepam}). Starting from a completely random state presents a challenge, particularly for larger systems. Ideally, the process should begin from a local minimum, which can be obtained through methods such as MD runs, CG relaxation, or ARTn coupled with a semi-empirical model. In this section, we present preliminary results that demonstrate the capability of ARTn-MTP to push the limits of relaxation for a configuration already optimized by ARTn coupled with the Stillinger-Weber (SW) potential \cite{stillinger1985computer}.

A previously published a-Si model \cite{barkema1996event, malek2000dynamics} using ARTn with the SW potential \cite{stillinger1985computer}, modified by Vink \cite{vink2001fitting}, served as input for an ARTn-MTP process. The 1000-atom ARTn-SW a-Si configuration (1000-aSi-SW initial) was refined through ARTn-MTP, resulting in the (1000-aSi-SW final) configuration. The goal was to assess the ARTn-MTP process's ability to refine configurations optimized with a semi-empirical potential. The energy dropped by -23.25 meV/atom, and structural properties, shown in Fig. \ref{fig:swconf} and summarized in Tab. \ref{tab:DefectsSW}, reveal significant improvements. The number of dangling and floating bonds decreased, reducing total defects by 3\%. Bond angles (Fig. \ref{fig:swconf} b) are closer to the ideal 109.5°, with a narrower spread in the final model. Bond lengths were concentrated around 2.38 Å, aligning with experimental values, and the radial distribution function in Fig. \ref{fig:swconf} c confirmed this. The 1000-aSi-SW final and 100k-GAP18 models showed better agreement with experimental data than the initial configuration. The structure factor, in Fig. \ref{fig:swconf} d, also matched experimental S(q). The excess energy (\(\Delta E\)) decreased from 0.2021 eV/atom (initial) to 0.1789 eV/atom (final), falling within the experimental range of 0.135 – 0.205 eV/atom \cite{limbu2018information}. The final configuration, with lower excess energy, is more stable and realistic. The bulk modulus increased from 65.8573 GPa (initial) to 68.8690 GPa (final), indicating enhanced stiffness. These results demonstrate that the ARTn-MTP coupling scheme effectively relaxes a semi-empirically optimized amorphous configuration.

\begin{table} 
\begin{tabular}{lcc}
\hline
 Properties& \makecell{1000-aSi-SW \\ initial} & \makecell{1000-aSi-SW \\ final}  \\
 \hline
Density (g/cm$^3$)   & 2.20  & 2.24 \\
Fourfold Si (\%)  & 94.60  & 97.60 \\
Threefold Si (\%)  & 3.00  & 0.90 \\
Fivefold Si (\%)  & 2.40  & 1.50 \\
Average bond lenght ($\sigma$) & 2.38 (0.05)  & 2.38 (0.03) \\
Average bond angle ($\sigma$) & 109.11 (10.91)  & 109.12 (11.26) \\ 
Average coordination number & 3.99  & 4.01 \\
$\Delta E$ (eV / atom)  &  0.20 &  0.18\\
Bulk modulus (GPa)  &  65.86 &  68.87 \\
Crystallinity (\%) &0.00 & 0.00\\
\hline
\end{tabular} 
\caption{\label{tab:DefectsSW} Structural and eleastic properties for 1000-aSi-SW initial and 1000-aSi-SW final  configurations: coordination defects, coordination number, average bond length, average bond angle, bulk modulus and energy difference per atom between amorphous state and crystalline Si. Standard deviations are provided in parentheses.  The ART-SW optimized configurations (1000-aSi-SW initial) served as the input for the ARTn-MTP refined configurations (1000-aSi-SW final).}
\end{table}

\subsection{Stiffness}
 \begin{figure}
\centering
    \includegraphics[width=8.5cm]{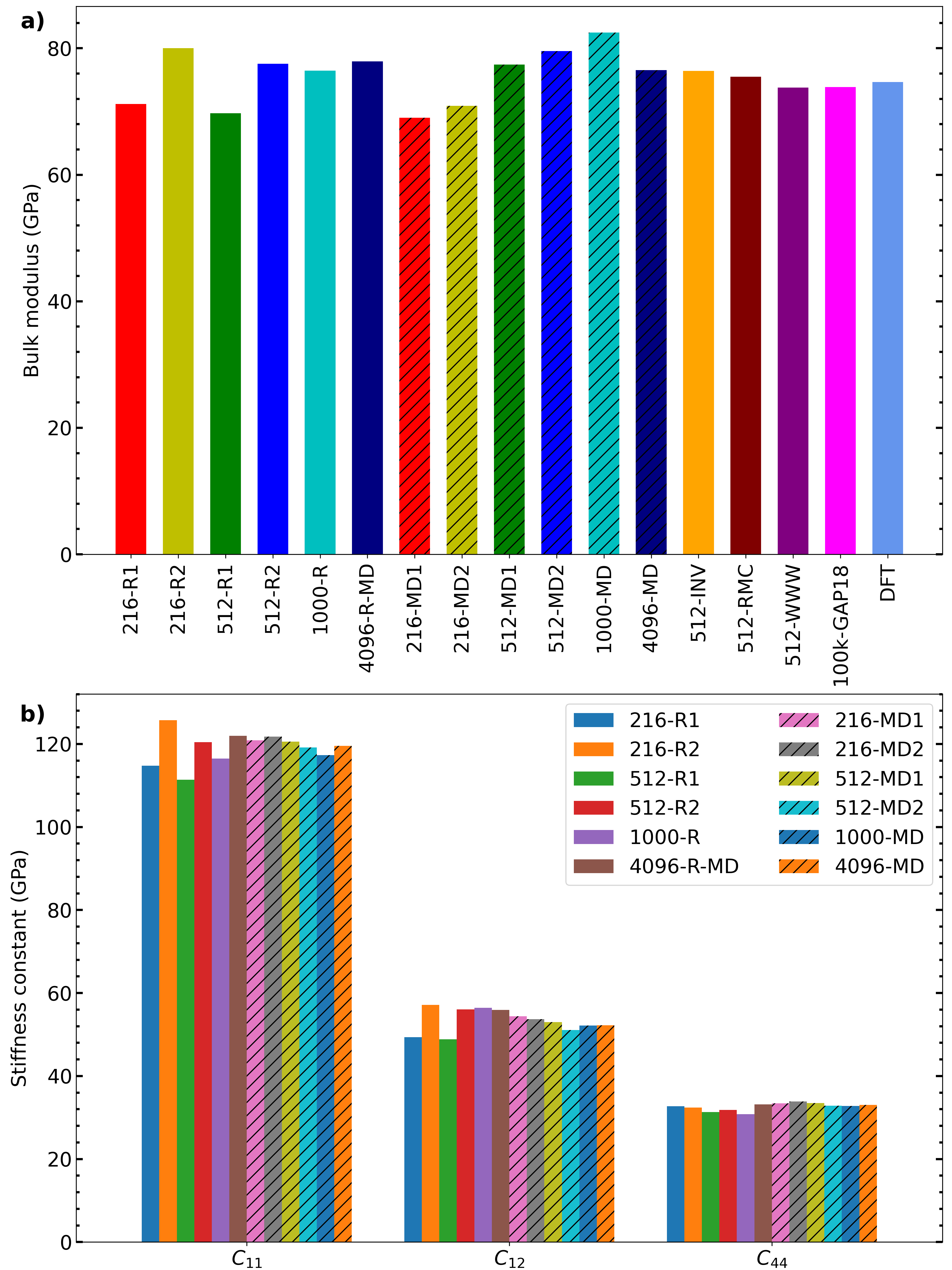}
    \caption{\label{fig:Bmo} Mechanical properties of the models: a) Bulk modulus of ARTn-MTP generated models, MD generated models (hatched bars), and literature models.b) Elastic constants of ARTn-MTP generated a-Si models and MD simulated models (hatched bars). }
\end{figure}

We calculated the bulk modulus \( B \) of our models and of the models found in the literature. The results are presented in Fig. \ref{fig:Bmo} a. We include the bulk modulus value obtained from density functional theory (DFT) simulations reported in Ref.~\cite{durandurdu2001ab}, which is 82.5 GPa. \ZK{The DFT value was computed using the local density approximation (LDA) and hard, norm-conserving pseudopotentials \cite{durandurdu2001ab}}. For our 216-R2 configuration with 100\% 4-fold coordination, we obtained a bulk modulus of 80.01 GPa. The 512-R2 model yields a bulk modulus of 77.53 GPa, closely matching the 512-WWW model's value of 77.41 GPa. In contrast, other models such as 512-MD, 512-INV, and 512-RMC exhibit lower bulk modulus values of 71.18 GPa, 69.00 GPa, and 70.90 GPa, respectively. The 1000-R model demonstrates a bulk modulus of 76.46 GPa. Our largest models, the 4096-R-MD and 100k-GAP18, yield bulk modulus values of 77.93 GPa and 79.56 GPa, respectively.\ZK{For our a-Si models generated solely using MD simulation, the bulk moduli are as follows: 76.54 GPa for 216-MD1, 76.41 GPa for 216-MD2, 75.51 GPa for 512-MD1, 73.77 GPa for 512-MD2, 73.88 GPa for 1000-MD, and 74.66 GPa for 4096-MD. With the exception of 216-MD2, which shows no crystallinity, the other MD models contain significant proportions of crystallinity. The bulk modulus values presented here reflect the quality of our amorphous models. It is important to note that the bulk modulus for amorphous materials depends heavily on the preparation method and the inherent lack of periodicity in the structure. As a result, the value of the bulk modulus is not unique and varies across different models. The average bulk modulus of all the models, including our ARTn models, MD models, and those from the literature, is 75.47 GPa, with a spread of 3.63 GPa. All the bulk modulus values, including the mean, are notably lower than the bulk modulus of crystalline silicon, which is 98.3 GPa \cite{brandes2013smithells}.  A value of 59 GPa for the experimental bulk modulus of hydrogenated amorphous silicon (a-Si:H) has also been reported \cite{tanaka1986elastic}. This value is significantly lower than the bulk modulus values observed in all of the models presented. All these values were obtained using our unified machine learning potential developed for the Si, O, and SiO$_2$ systems. Other potentials may yield different results. To our knowledge, there are no clear experimental benchmark values for the bulk modulus of amorphous silicon. In addition, we also assessed the elastic constants of all our models, and the results are presented in Fig. \ref{fig:Bmo} b. As can be noted, strong dispersion is not noticeable. For our ARTn-MPT models, the mean values (with spreads) are as follows: 118.46 GPa (4.77 GPa) for C$_{11}$, 53.99 GPa (0.00 GPa) for C$_{12}$, and 32.05 GPa (0.79 GPa) for C$_{44}$. For our a-Si models generated solely with MD, the corresponding values are 119.87 GPa (1.45 GPa) for C$_{11}$, 52.76 GPa (1.08 GPa) for C$_{12}$, and 33.24 GPa (0.38 GPa) for C$_{44}$. These values lie within the same range as those reported in previous works: 121 GPa for C$_{11}$, 43 GPa for C$_{12}$, 40 GPa for C$_{44}$, and 69 GPa for Bulk modulus \cite{deringer2018realistic}. Thus, while mechanical properties are highly dependent on the quality of the potential, we can argue that the computed values presented here fall within the range of the true mechanical properties of real a-Si models.}

\subsection{Potential energy} 
\begin{figure}
\centering
    \includegraphics[width=8cm]{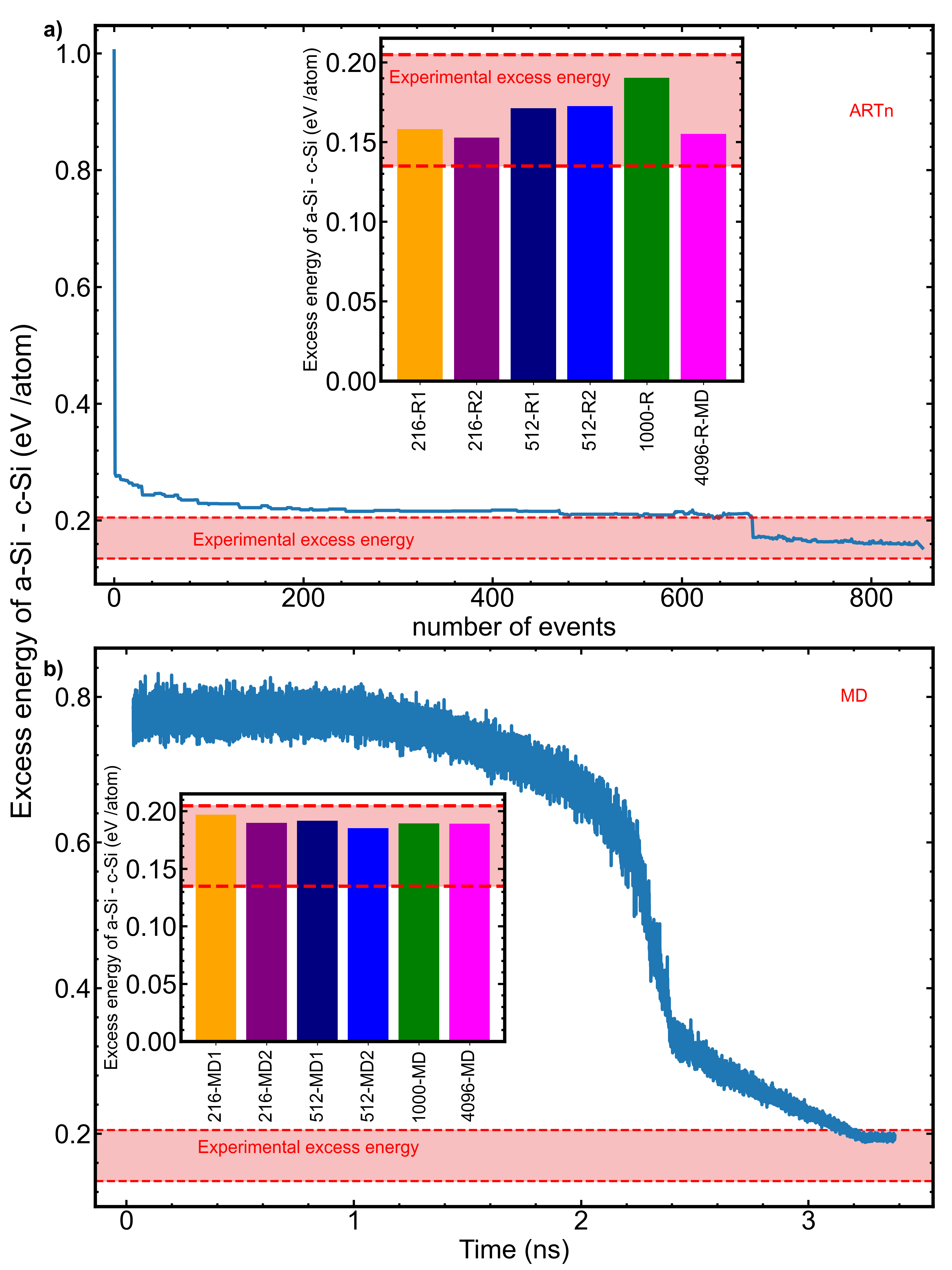}
    \caption{\label{fig:edeac} a) Excess energy between ART-MTP configurations (a-Si) and crystalline silicon (c-Si) as a function of the number of accepted events. A total of 4 events per atom was required to achieve 100\% 4-fold coordination in 216-R2 a-Si model. b) Excess energy for models generated using MD simulation solely}
\end{figure} 
\begin{table}

\begin{tabular}{lccc}
\hline
Models & $\rho_i$ (g/cm$^3$) & $\Delta E$ (eV / atom) & $\rho_f$ (g/cm$^3$) \\
 \hline
216-R1  & 2.23 & 0.16 & 2.24 (2.23) \\
216-R2  & 2.28 & 0.15 & 2.25 (2.24) \\
512-R1  & 2.23 & 0.18 & 2.23 (2.22)  \\
512-R2  & 2.28 & 0.18 & 2.25 (2.24)  \\
1000-R   & 2.28&  0.19 & 2.25 (2.24) \\
4096-R-MD  & 2.28 & 0.16 & 2.26 (2.25) \\
Exp  & 2.2800 \cite{custer1994density}  &  0.135 – 0.205 \cite{limbu2018information} & - \\
\hline
\end{tabular} 
\caption{\label{tab:Density} Initial ($\rho_i$) and final ($\rho_f$) density for the ARTn-MTP generated a-Si model. Each model was randomly filled and the simulation box sizes was adjusted to match the initial density, which was maintained from the beginning to the end of the ARTn simulation.  The final density was obtained by relaxing each of the generated models to 0.5 K as described in main test. The values in parentheses were obtained by relaxing each model at 300 K and 1 atm. The excess energy  in eV per atom ($\Delta E$) relative to crystalline silicon is presented alongside the experimental excess energy.}
\end{table}

Fig. \ref{fig:edeac} a shows the relationship between energy decrease and the number of events for the model 216-R2. As can be seen, the energy decreases as the number of events increase. A detailed analysis of the events and key parameters is provided in the Nature of the Relaxation section. The experimental excess energy interval is indicated by the light red shaded region in the figure for reference. The inset of Fig. \ref{fig:edeac} a shows the excess energy per atom for each of our models. These values are also presented in table \ref{tab:Density}. 
All models fall within the experimental excess energy interval \cite{limbu2018information}, as detailed in the inset. Notably, the 216-R2 model, which initially contained over 70\% coordination defects, was efficiently relaxed to achieve 100\% 4-fold coordination and the lowest excess energy of 0.17 eV/atom with just 4 events per atom. To our knowledge, with the exception of a few Keating-potential-based WWW models \cite{wooten1985computer,pandey2016inversion,barkema2000high}, no other method, including DFT \cite{pedersen2017optimal}, has yielded a model of a-Si with no coordination defects. \ZK{Fig. \ref{fig:edeac} b shows the evolution of excess energy as a function of simulation time, with the inset corresponding to the excess energy of the generated a-Si models. Once again, the excess energy falls within the experimental range, although the values are higher than those for the ARTn-MTP-generated a-Si models. The mean excess energy for the ARTn-MTP models is 0.1667 eV per atom, with a spread of 0.0130 eV per atom, while for the MD-generated models, the mean excess energy is 0.1905 eV per atom, with a spread of 0.0035 eV per atom. This demonstrates that the ARTn-MTP-generated a-Si models are more stable than those generated by MD-MTP.}

\subsection{Density} 
We computed the density of our models at 0 K. The final density was obtained by running a 50 ps molecular dynamics simulation, cooling the system from 300 K to 0.5 K, and then holding it at 0.5 K for an additional 50 ps, while barostatting the system at 0 Pa. In separate simulations, we also equilibrated each of the aforementioned models at 300 K and 1 atm for 100 ps. The results are presented in table \ref{tab:Density}. Recall that \ZK{the calculated density and experimental density of crystalline Si are 2.28 g/cm³ (variable cell relaxation 0 K) and 2.33 g/cm³ (300 K), respectively. The calculated density was obtained using an MTP potential that was trained on DFT data ( 0 K calculations). Therefore, a reasonable comparison between the two densities could not be made due to the difference in temperature conditions. We recorded a mean deviation of 1.46\% from the calculated value and 3.57\% from the experimental value of crystalline silicon.} The MTP potential is trained using a generalized gradient approximation exchange-correlation functional, which is known to overestimate the density of crystalline silicon by 2 \% \cite{jain2015effect}. Retraining the MTP on a higher-accuracy exchange-correlation functional, such as the restored regularized strongly constrained and appropriately normed meta-generalized gradient approximation \cite{furness2020accurate} is likely to solve this small discrepancy with experimental values. 

\subsection{Nature of the relaxation}
\begin{figure}
\centering
    \includegraphics[width=8.3cm]{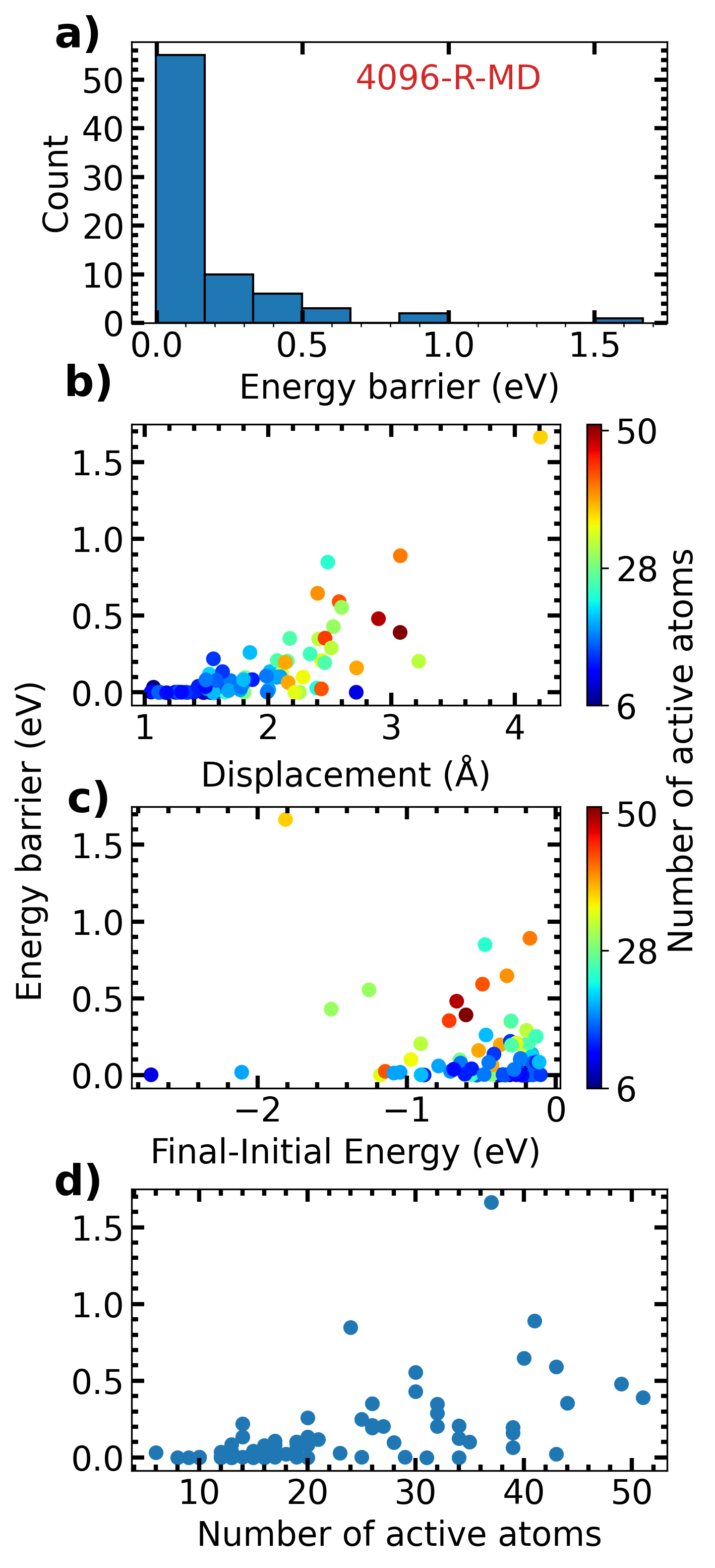}
    \caption{\label{fig:Barriers} Analysis of accepted Events in ARTn-MTP optimization: a) Barrier distribution, b) square root of squared atomic displacement versus energy barrier, c) energy barrier as a function of energy decrease, and d) energy barrier versus active atoms during events}
\end{figure}

The primary goal is not only to generate high-quality amorphous structures but also to understand the underlying physics of the relaxation process in amorphous systems over time. Thus, to understand the relaxation process in our amorphous system, we analyze simulation data, focusing on key parameters such as barrier distribution, atomic displacements, active atoms, and energy decrease, considering only accepted events as detailed in Fig. \ref{fig:Barriers}. These parameters can help provide insights into the nature and mechanism of the relaxation process. The data were collected under normal deformation, excluding shear deformation during the simulation. We first examine the correlation between atomic displacement and energy barriers. A clear correlation is observed: larger displacements are typically associated with higher energy barriers and more active atoms, as shown in Fig.~\ref{fig:Barriers} a. To further validate this, we calculate the correlation between the square of the atomic displacement and the energy barriers. According to Hooke's law, we expect a quadratic relationship between energy and displacement. When we plot energy against the square of the displacement, we expect a linear relationship, which our results confirm, yielding a strong correlation. \ZK{A fit was performed to identify potential trends linking atomic displacement to the system's energy. For instance, Hooke's law is often applied in similar contexts, where atomic displacement within a certain range is used to determine the spring constant. In this case, a log-log fit indicates a power law exponent of 2.4 with a standard error of ±0.7, considering only the ARTn events in the 4,096-atom system. An analysis over all the ARTn events generated in this study reveals a power law exponent of 1.90 with a standard error of ±0.04. These results suggest that Hooke's Law likely applies.} Next, we find no significant correlation between the energy barriers and the energy gain, nor between energy gain and the number of active atoms, which is consistent with previous studies \cite{kallel2010evolution}. Furthermore, no correlation exists between the number of atoms displaced up to the saddle point and the final state. 
As illustrated in Fig. \ref{fig:Barriers} d, a correlation exists between energy barriers and the number of active atoms, with higher energy barriers typically involving a greater number of active atoms. On average, the number of active atoms varies from 10 to 30, with a mean of around 20 atoms that moves by more than 0.1~Å per event. This suggests that each event is localized, typically involving a cutoff radius of less than 7 Å. The microstructural changes primarily occur within a few neighboring cells, indicating a localized relaxation process. The relationship between the aforementioned parameters and the number of randomly displaced atoms at the start of ARTn searches (which depends on the cutoff and the central atom) was not investigated in this works.

\subsection{Computational cost and nonlinear optimization process}
\begin{figure}
\centering
    \includegraphics[width=8.5cm]{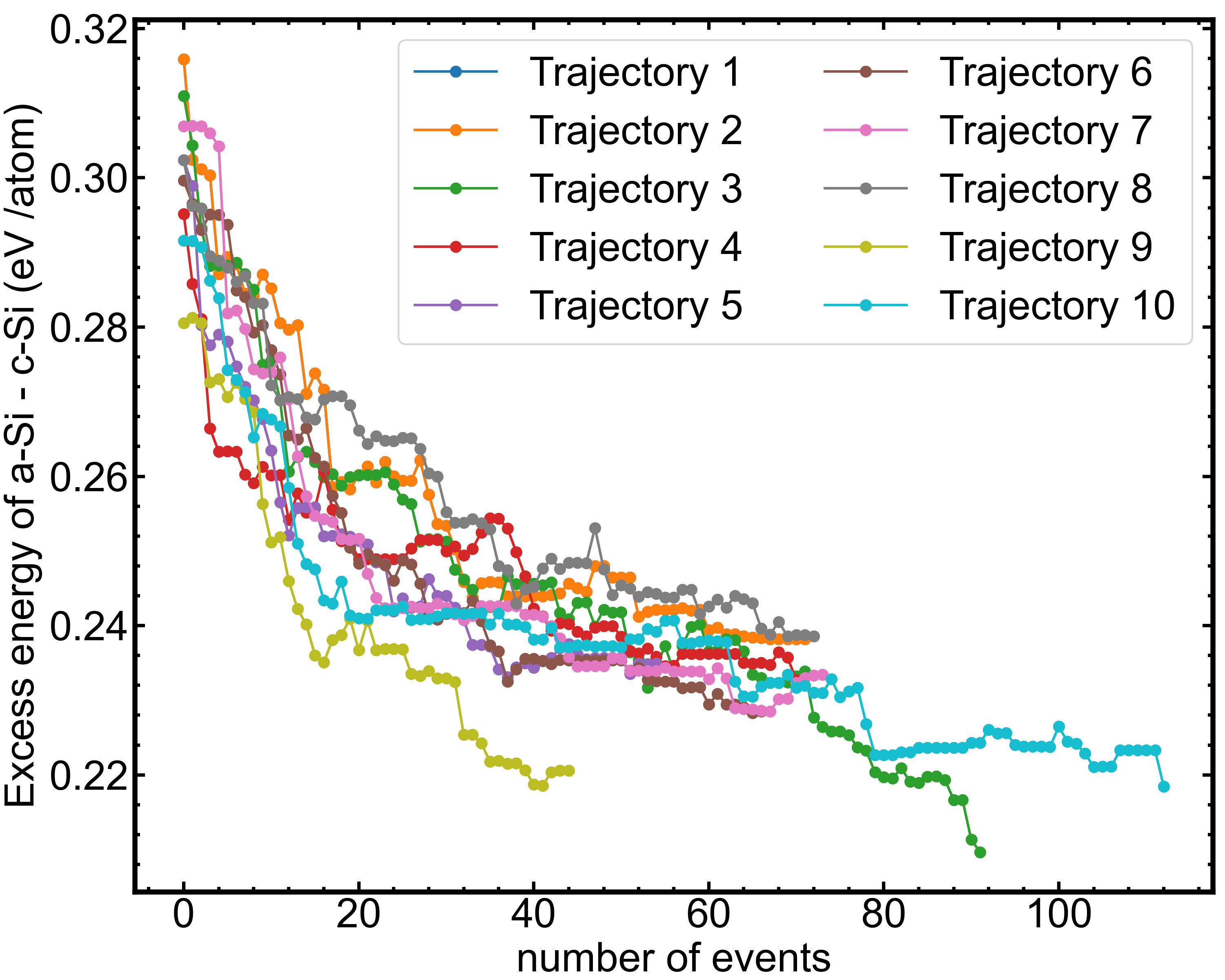}
    \caption{\label{fig:nonlinear} Nonlinear optimization effects characterized by excess energy per atom and number of events computed from independent runs with identical input Configuration}
\end{figure}

One of the primary challenges in computational materials modeling is finding the right balance between accuracy and computational cost. In this work, we present the first-ever coupling between ARTn and machine learning potentials, and we document the computational cost involved in preparing our amorphous system.

For the system of 216 atoms, the optimization of configurations required 0.0615 core-years and involved 55.20 millions force evaluations. Notably, one of the 216 atoms was optimized to achieve 100\% 4-fold coordination. For the larger systems, the computational costs were as follows: the 512-atom system took 0.22 core-years, and the 1000-atom system required 0.34 core-years. Finally, for the 4096-atom system, we recorded 0.09 core-years to refine the MD configuration, with 37.25 million force evaluations. As mentioned in previous sections, this configuration achieved 98.05\% 4-fold coordination and an excess energy of 0.15 eV per atom, which is closer to the experimentally observed low excess energy value.

The nonlinear optimization process within the ARTn framework was explored using a randomly generated 216-atom input configuration, with independent runs conducted using the same number of steps. Data were collected from these independent trajectories. Fig. \ref{fig:nonlinear} illustrates the relationship between excess energy levels and the number of accepted events for each trajectory. The results reveal varying numbers of accepted events and relaxation levels across the trajectories. Among the independent runs, the lowest number of accepted events was 45, while the highest was 113. Despite this difference, both systems achieved nearly identical excess energy levels. The mean number of accepted events was approximately 75, with most systems exhibiting higher excess energy than the system with 92 accepted events, which showed the lowest excess energy. These findings are indicative of a nonlinear optimization process.

\section{Conclusions}

\ZK{This paper investigates a-Si using ARTn coupled with a MTP. Seven models of different sizes were developed, showing excellent agreement with both experimental data and a 100,000-atom MD-based model using a different MLIP. The ARTn-MTP optimizer refines configurations optimized by ARTn with a semi-empirical model. It also improves the relaxation of a-Si generated through the melt-quench process. High-quality amorphous structures with 216, 512, and 1000 atoms were produced, all free of local crystallinity. Starting ARTn simulations with a randomly filled configuration yields less crystallinity compared to using an MD-prerelaxed configuration. Overall, ARTn-MTP outperforms MD-MTP in modeling a-Si, as MD-generated a-Si often shows significant crystallinity, whereas ARTn-MTP produces a-Si with minimal crystallinity and low coordination defects. }

\ZK{The relationship between local crystallinity, defects, and ideal amorphous structures remains underexplored due to a lack of high-quality models from different pathways like ARTn and melt-quench. Our results highlight new research directions for improving a-Si understanding and fabrication. Future work could focus on increasing system sizes while minimizing defects by targeting over- and under-coordinated atoms. Additional studies will investigate crystallinity’s origins, its formation in a-Si, and methods for producing large-scale amorphous a-Si without crystallinity.} 

\ZK{While ARTn combined with MLIPs is effective for modeling a-Si, further innovation is needed to replicate experimental conditions more accurately. a-Si is mainly produced by chemical vapor deposition and ion implantation, with representative volumes ranging from tens of nanometers to micrometers and processing times from seconds to hours. Achieving these time- and length-scales is essential for both applied and fundamental studies. Despite advancements in parallel computing, reaching long timescales remains a challenge.} 

\section*{Data availability}

The potential file and the a-Si models are publicly available at \url{https://gitlab.com/Kazongogit/MTPu/-/tree/main/ART/Si?ref_type=heads}. The MTP source code can be found at \url{https://mlip.skoltech.ru/}, and the ARTn source code is freely available through a request to Normand Mousseau.

\bibliographystyle{unsrt}
\bibliography{ref1}
\section*{Acknowledgements}
\noindent We thank the Digital Research Alliance of Canada for generous allocation of compute resources. Financial support was provided by the Natural Sciences and Engineering Research Council of Canada (NSERC) and the Association canadienne-française pour l'avancement des sciences (ACFAS), and the Canada Research Chair on Sustainable Multifunctional Construction Materials.
\section*{Author contributions}
\noindent L.K.B. initiated and coordinated the research project. K.Z. performed all ARTn-MTP and molecular dynamics simulations, as well as the analyses presented in this work. All authors—K.Z., H.S., N.M., L.K.B., and C.O.P.—contributed to the writing of the paper. C.O.P., N.M., and L.K.B. supervised K.Z., while C.O.P. and L.K.B. provided the necessary computing resources.
\section*{Competing interests}
\noindent The authors declare no conflict of interest.
\end{document}